%% file: paper.tex
\documentclass[12pt,letterpaper]{article}

\usepackage{fullpage}
\usepackage[utf8]{inputenc}
\usepackage[russian]{babel}
\usepackage{amsmath}
\usepackage{amssymb}
\usepackage{amsthm}

\title{Однородная память и сериализация для ${\lambda\beta\eta}$}
\author{Салихметов~А.~Е.}

\newtheorem{theorem}{Утверждение}
\theoremstyle{definition}
\newtheorem{definition}{Определение}

\begin{document}

\maketitle

\begin{abstract}
В данной статье мы введем особый вид систем, определим их
свойства, а затем покажем, что машину редукции на основе
экстенсионального бестипового $\lambda$-исчисления можно
реализовать как одну из таких систем.
Более точно, речь пойдет об однородной памяти и работе в
реальном времени, а также о частном случае сериализации,
которая пригодится нам для механизма сравнения результатов
прямо во время вычисления.
Начнем мы, тем не менее, с того, какова мотивация работы
над данной темой, а также какое место занимает сама теория
${\lambda\beta\eta}$ в аналитическом круговороте автора.
\end{abstract}

\input intro
\input constr
\input subject
\input emu
\input lambda
\input complete
\input recreal
\input serial
\input dict
\input code
\input nequal
\input ext
\input unlimnf
\input btio
\input uniform
\input property
\input strategy
\input mlc
\input sugar
\input overview
\input plan
\input first
\input second
\input third
\input end
\input appendix

\input biblio
\end{document}

%% file: intro.tex
\section{Введение}

История $\lambda$-исчисления, которая уходит в начало
прошлого века, подробно освещена в \cite{history}.
Там же затрагивается такая малоизвестная, но любопытная деталь,
как этимология названия данного раздела математической логики,
который служит основой для <<computer science>>.
Сам значек <<$\lambda$>> используется для одной из двух основных
конструкций в созданной Черчем системе~--- абстракции.
Оказывается, что выбор обозначения абстракции не был совершенно
случайным, а сделан в противопоставление другой более ранней
конструкции, которую использовали Whitehead и Russell и
обозначали как <<$\hat x$>>.
Для новой конструкции, чтобы отличать ее от прежней, Черч сначала
заменил обозначение сначала на «${\land x}$», а затем~--- на
«${\lambda x}$», очевидно, интерпретировав первый символ как
заглавную букву «$\Lambda$», для упрощения набора.

Вычисление, основанное на теории $\lambda$-исчисления, было
реализовано во многих языках программирования \cite{sicp},
а также некоторых машинах редукции, в том числе~--- машинах
редукции графов с ленивым связыванием подвыражений.
Последняя конструкция принадлежит Wadsworth \cite{graphred}.
С различными аспектами реализации аппликативных вычислений
читатель может ознакомиться, например, по \cite{kluge} и
\cite{funcprog}.

В рамках данной статьи нас будут особенно интересовать
<<explicit substitution>> и <<term rewriting systems>>.
Сконцентриуем мы внимание на системе аксиом, введенных в
\cite{revesz} и затрагиваемых позднее в различных источниках,
включающих \cite{ustica}, где замена $\beta$-редукции
четырьмя правилами была названа <<Micro Lambda Calculus>>.
Позднее в личном сообщении \cite{private} Vincent van Oostrom
предложил термин <<дистрибутивное $\lambda$-исчисление>>, пытаясь
снабдить полученную систему более формальным наименованием.
Данная система интересна тем, что позволяет описать <<explicit
substitution>>, не вводя никаких дополнительных обозначений
поверх чистого бестипового $\lambda$-исчисления, в отличие
от остальных подходов (см., например, \cite{explicit}).
Упомянутое личное сообщение будет изложено в приложении ради
положительного закрытия прежде открытого вопроса о существовании
эффективной (рекурсивной) одношаговой стратегии для
дистрибутивного $\lambda$-исчисления (нормальная стратегия
нормализующей для него не является).

Приведем краткий план изложения материала в настоящей работе.
Начнем мы с того, какое место занимает $\lambda$-исчисление
вообще и система ${\lambda\beta\eta}$ в частности (читателя ждет
краткое определение этой формальной теории в одном из первых
разделов) в аналитическом круговороте автора; тем самым мы
обозначим мотивацию работы над данной темой.
Затем мы определим специальный способ сериализации
$\lambda$-выражений префиксными бинарными кодами
(обычно рассмотрение сериализации ограничивается лишь так
называемыми Геделевскими номерами \cite{monograph}).
Наконец, мы введем особый класс систем на однородной памяти
(это понятие будет строго определено позднее) и их свойства
и покажем, что их достаточно для реализации машин редукции
графов, а также приведем конструктивный пример одной из них.

Основной целью, преследуемой в настоящей работе, является
построение моста между <<The Heap Lambda Machine>> из \cite{heap}
и проектом <<Macro Lambda Calculus>> с подведением более
систематической теоретической основы для обоих систем.

%% file: constr.tex
\section{Счетность в конструктивизме}

Исходные объекты для построения теорий всегда выбраны из счетного
множества: например, из имен; имена счетны ввиду счетности языка
как множества текстов.
Далее, при попытке ввести множество всех подмножеств такого или
некоторого производного счетного множества возникает проблема
континуума.
Однако, конструктивно доказать, что континуальные множества
вообще существуют, не представляется возможным.

Проблема с континуальными множествами довольно остро проявилась
в прошлом веке при развитии теории вычислимости,
где требовалась математическая модель для счетных множеств,
которые бы состояли из отображений внутри него.
Разрешилась она тогда, когда Скотт предложил топологию дерева и
урезал отображения до непрерывных.
С другой стороны, привести конструктивный пример функции, которая
вне такого множества, невозможно, так как язык описывает
в точности частично-рекурсивные функции.

Возникает логичный вопрос о том, как же быть с вещественными
числами, и до какого множества следует их урезать, чтобы описать
именно те, которые можно определить конструктивно.
Такие числа есть в конструктивном вещественном анализе, и
называются рекурсивными вещественными числами.

В принципе, на каждом этапе, где требуются новые конструкции,
в традиционной математике из-за неконструктивных определений
обычно ведущие к несчетным множествам, можно каждый раз вводить
надлежащую топологию и урезать отображения до непрерывных.
Утверждается, что это возможно для любых мыслимых конструкций,
что таким образом освобождает от интуитивно присутствующей
опасности ограниченности конструктивного подхода.

Наконец, учитывая, что система ${\lambda\beta\eta}$ соответствует
максимальным непротиворечивым теориям и при этом ввиду тезиза
Черча--Тьюринга имеет максимальную выразительность, любые
мыслимые конструкции оказываются возможными для определения
внутри нее.
И это может, в свою очередь, служить дополнительным аргументом
в сторону ее использования и большей концентрации внимания
на эквивалентных системах и на теории вычислимости вообще.

%% file: subject.tex
\section{Субъективное в языке как открытые термы}

Если математикой считать множество математических текстов $M$,
написанных с использованием одного из человеческих языков $H$,
то $M$ есть подмножество $H$.
Заметим, что если мы пронумеруем значения слов языка,
например, натуральными числами или эквивалентными
конструктивными объектами, то математическая часть может
описывать весь остальной человеческий язык.

Теперь, если мы ограничим рассмотренное подмножество
до конструктивной математики, то математическое подмножество
продолжит порождать человеческий язык, так как множество
текстов счетно.
Заметим, что в системе ${\lambda\beta\eta}$ описываются любые
конструктивные объекты, так как для этого необходимо и достаточно
иметь вычислимые функции, рекурсивно перечисляющие множества тех
или иных конструктивных объектов.
Будем считать, что человеческий язык построен на основе
${\lambda\beta\eta}$, не вдаваясь в детали или механизм
перечисления терминов; обратим внимание лишь на следующее.

У слов, которые тем или иным образом оказались закодированы
термами, есть некоторые параметры~--- свободные переменные~---
они должны обозначать все, от чего зависит объективная
интерпретация выражений-фраз, например, контекст, область,
ссылки.
Снова опустим детали и перейдем от слов, которые имеют значения,
необходимые для объективной (точной) интерпретации фразы,
к словам, которые таких значений не имеют (быть может, кроме как
в очень узких областях человеческой деятельности).
Например, слово <<вкусный>>~--- его свободные переменные,
по крайней мере, включают субъект, которому принадлежит данное
отношения к объекту.
Таким образом, мы можем, даже из слов, не имеющих объективного
(точного) значения~--- мы их будем называть субъективными
терминами~--- построить объективную фразу: абстрагировав субъект,
введя по сути общность, и сделав фразу неизбежно неверной
(так как всегда найдется субъект, для которого такое отношение
не имеет места), либо применив абстракцию к конкретному субъекту,
например, говорящему, и тогда анализировать фразу
не лишено смысла.
Разумеется, время есть тоже часть параметров фразы, но зачастую
во фразах время абстрагируется.

%% file: emu.tex
\section{Эмуляция ${\lambda\beta\eta}$ внутри нее самой}

Известно, что Тьюринг-полные системы отличаются тем, что
каждая из них может \textit{эмулировать} любую другую или
даже ее саму.
То есть из Тьюринг-полноты ${\lambda\beta\eta}$ следует
тот факт, что ее можно каким-то образом представить внутри
нее самой.
Однако показать, как именно построить такую эмуляцию и
эффективно ее реализовать, оказывается не самой легкой задачей.
Более формально она выражается в двух следующих пунктах.

\begin{enumerate}
\item Определение функции-вычислителя нормальной формы выражений
в виде комбинатора.
\item Определение представления произвольных выражений в форме,
подходящей для вычисления с помощью функции, определяемой в (1).
\end{enumerate}

Очевидно, что существует бесконечное счетное множество
тривиальных решений данной задачи, сводящихся к следующему:
определить тождественные и функцию-вычислитель, и преобразование
выражений во внутреннюю форму.

Но также была найдена часть нетривиального решения:
в \cite{selfinter} вводится комбинатор для нахождения
$\beta$-нормальной формы $\lambda$-выражений, представляемых
через определенное кодирование, внутри самого
$\lambda$-исчисления.

%% file: lambda.tex
\section{Краткое определение ${\lambda\beta\eta}$}

Без лишних комментариев введем множество $\lambda$-термов
$\Lambda$, которое состоит из переменных $x$, $y$, $z$ и~т.~д.,
с которыми мы будем здесь работать без особых осторожностей,
абстракций ${\lambda x. M}$ и аппликаций ${M\, N}$, затем
подстановку ${M[x := N]}$, множество свободных переменных
${\mathrm{FV}(M)}$; заявим $\alpha$-конвертируемые термы
текстуально равными; определим редукцию ${\beta\eta}$,
рефлексивное и транзитивное замыкание отношения <<быть
подвыражением>>, и, наконец, результат одного шага редукции
${F_l(M)}$ при нормальной стратегии вычисления.
За более аккуратным определением читатель всегда может обратиться
к \cite{monograph}.
\begin{eqnarray*}
x, y, z, \dots &\in& \Lambda; \\
M, N \in \Lambda &\Rightarrow& \lambda x. M \in \Lambda
\land M\, N \in \Lambda; \\
(M) \equiv M &\land& M\, N\, P \equiv (M\, N)\, P.
\end{eqnarray*}
\begin{eqnarray*}
x[x := P] &\equiv& P; \\
y[x := P] &\equiv& y; \\
(\lambda x. M)[x := P] &\equiv& \lambda x. M; \\
(\lambda y. M)[x := P] &\equiv& \lambda y. M[x := P]; \\
(M\, N)[x := P] &\equiv& M[x := P]\, N[x := P].
\end{eqnarray*}
\begin{eqnarray*}
\mathrm{FV}(x) &=& \{x\}; \\
\mathrm{FV}(\lambda x. M) &=& \mathrm{FV}(M) \setminus \{x\}; \\
\mathrm{FV}(M\, N) &=& \mathrm{FV}(M) \cup \mathrm{FV}(N).
\end{eqnarray*}
\begin{eqnarray*}
y \not\in \mathrm{FV}(M) &\Rightarrow&
\lambda x. M \equiv \lambda y. M[x := y].
\end{eqnarray*}
\begin{eqnarray*}
\beta &=& \bigl\{((\lambda x. M)\, N, M[x := N]) \bigm|
M, N \in \Lambda\bigr\}; \\
\eta &=& \bigl\{(\lambda x. M\, x, M) \bigm|
M \in \Lambda, x \not\in \mathrm{FV}(M)\bigr\}; \\
\beta\eta &=& \beta \cup \eta.
\end{eqnarray*}
\begin{eqnarray*}
M \subset M \subset &\lambda x. M; & \\
M \subset &M\, N& \supset N; \\
M \subset N \land N \subset P &\Rightarrow& M \subset P.
\end{eqnarray*}
\begin{eqnarray*}
&F_l(x)& \equiv x; \\
\exists Q \in \Lambda: (M, Q) \in \beta\eta \Rightarrow
&F_l(M)& \equiv Q; \\
\nexists Q \in \Lambda: (\lambda x. M, Q) \in \beta\eta
\Rightarrow &F_l(\lambda x. M)& \equiv \lambda x. F_l(M); \\
\nexists Q \in \Lambda: (M\, N, Q) \in \beta\eta \land
\exists P \subset M: \exists Q \in \Lambda: (P, Q) \in \beta\eta
\Rightarrow &F_l(M\, N)& \equiv F_l(M)\, N; \\
\nexists Q \in \Lambda: (M\, N, Q) \in \beta\eta \land
\nexists P \subset M: \exists Q \in \Lambda:
(P, Q) \in \beta\eta \Rightarrow &F_l(M\, N)& \equiv M\, F_l(N).
\end{eqnarray*}

%% file: complete.tex
\section{Полнота экстенсионального $\lambda$-исчисления}

Приведем обрывочную цитату из \cite{monograph} о полноте
${\lambda\beta\eta}$:
<<Так как $\lambda$-теория свободна от логики, непротиворечивость
нужно понимать следующим образом...
Равенство~--- это формула вида ${M = N}$, где $M$, $N$~---
$\lambda$-термы; такое равенство замкнуто, если $M$, $N$
не содержат свободных переменных...
Пусть $T$~--- формальная теория, формулами которой являются
равенства.
Тогда говорят, что $T$ непротиворечива (и пишут
${\mathrm{Con}(T)}$), если в $T$ доказуемо не любое замкнутое
равенство.
В противном случае говорят, что $T$ противоречива...
Одна из причин рассмотрения системы ${\lambda\eta}$ состоит
в том, что оно обладает определенным свойством полноты...
Эквациональная теория $T$ называется полной по Гильберту--Посту
(сокращенно HP-полной), если для любого равенства ${M = N}$
в языке теории $T$ или ${M = N}$ доказуемо, или ${T + (M = N)}$
противоречива...
HP-полные теории соответствуют максимальным непротиворечивым
теориям в теории моделей для логики первого порядка>>.

%% file: recreal.tex
\section{Рекурсивные вещественные числа}

Рассмотрим множество всех комбинаторов, определенных на
нумералах, которые каждому нумералу сопоставляют список вида
${[b, \lceil m \rceil, \lceil n \rceil]}$, где $m$ и $n$~---
натуральные числа, а $b$~--- булево значение ($T$ или $F$),
которое представляет знак (<<$+$>> или <<$-$>>).
Тогда список соответствует рациональному числу ${\pm m / n}$.

Теперь, если $S_i$ есть рациональное число, соответствующее
${M\, \lceil i \rceil}$, $M$ из рассматриваемого множества
комбинаторов соответствует
${\lim_{n \rightarrow \infty} \sum_{i = 0}^n S_i}$.
Такое множество замкнуто относительно операции взятия предела
последовательности, как и множество вещественных чисел, но
не существует конструктивного примера вещественного числа,
которое не принадлежит построенному множеству.

Только что введенное множество чисел-комбинаторов эквивалентно
так называемым <<рекурсивным вещественным числам>>, определяемым
в \cite{recreal}.

%% file: serial.tex
\section{Сериализация $\lambda$-выражений}

Известно, что любое $\lambda$-выражение можно детерминированно,
хотя и не в реальном времени, перевести в комбинаторную логику
с одноточечным базисом.
Выражения в последней эквивалентны скобочным структурам, так как
являются, по сути, бинарными деревьями без нагрузки в узлах.
Такие структуры можно представить префиксными бинарными кодами
переменной длины, если заменить символ <<(>> битом $1$,
а <<)>>~--- битом $0$, а также добавить завершающий бит $0$.
Полученные коды можно передавать по сети и сохранять в файлы,
то есть использовать данный метод для сериализации
$\lambda$-выражений, если считать выражения равными с точностью
до ${\alpha\beta\eta}$-конверсии.
Последнее~--- существенное ограничение данного подхода, так как
десериализация в общем случае не восстанавливает исходное
$\lambda$-выражение текстуально, даже с точностью до
$\alpha$-конверсии.

С другой стороны, в \cite{strongred} приводится, по сути,
доказательство невозможности существования сильной редукции для
${\mathrm{CL}}$-теории, а точнее, доказывается, что набор правил
сильной редукции всегда бесконечен.
Такое свойство, следовательно, оказывается и у комбинаторной
логики с одноточечным базисом.

Это делает неприемлемым создание машины для сильной редукции на
основе комбинаторной логики.
Из известных теорий подходит для этой задачи лишь бестиповое
$\lambda$-исчисление с ${\beta\eta}$-редукцией, гарантирующее
получение по четкому алгоритму (при использовании одной из
нормализующих стратегий вычисления, например нормальной) так
называемой ${\beta\eta}$-нормальной формы, совпадающей для любых
двух эктенсионально равных выражений (выражения, не имеющие н.~ф.,
не считаются равными в ${\lambda\beta\eta}$, но есть вариант
модели, где считаются равными термы, не имеющие г.~н.~ф.).

%% file: dict.tex
\section{Словари $\lambda$-выражений}

Несмотря на то, что сериализация $\lambda$-выражений имеет
существенное ограничение: десериализация в общем случае
происходит с точностью до ${\alpha\beta\eta}$-конверсии,~---
данный способ может оказаться полезным для хранения
$\lambda$-выражений, их нормальных форм, кодов нормальных форм
и/или частично упрощенных вариантов, а также любой другой
связанной информации в словарях~--- словарях нормальных форм.

Это возможно потому, что коды, полученные в результате
сериализации любых двух $\lambda$-выражений, равных с точностью
до $\alpha$-конверсии, совпадают.
Таким образом можно получить словари нормальных форм столь скоро,
как получен способ приводить любые два экстенсионально равных
$\lambda$-терма к форме, совпадающей с точностью до
$\alpha$-конверсии.
Система, имеющая такое свойство, обязательно полна как
по Тьюрингу, так и по Гильберту--Посту.
Такая система может быть реализована, например, как было
запланировано ранее.

Следует отметить, что в самой системе ${\lambda\beta\eta}$
результат прямой десериализации ${\beta\eta}$-нормальных форм
в общем случае не является нормальной формой.
Однако, он всегда оказывается на одинаковом расстоянии от таковой
в смысле количества необходимых ${\beta\eta}$-редукций для всех
экстенсионально равных термов.

Подобные словари могут быть использованы для некоторых методов
вычисления $\lambda$-выражений, их представления
в <<человекочитаемом>> виде, хранения задач и их решений
в базах данных и т.~п.

%% file: code.tex
\section{Префиксное кодирование}

Сериализация $\lambda$-выражений, одним из возможных применений
которой ранее были рассмотрены словари н. ф. (нормальных форм),
может выполняться по ходу вычисления после получения г. н. ф.
(головной н. ф.), что, в свою очередь, может быть использовано,
в частности, для особого вида сравнения двух термов на предмет
равенства с точностью до ${\alpha\beta}$-конверсии
(экстенсиональность, в общем случае, не имеет места, так как,
например, терм ${\lambda x y. x\, (K\, I\, y)\, y}$ подлежит
$\eta$-редукции уже после получения г. н. ф.).

Так как предложенная сериализация предполагает перевод
$\lambda$-выражений в комбинаторную логику ${\{S, K, I\}}$,
а затем~--- в одноточечный базис ${\{X\}}$, то необходимо
показать, что для любой г. н. ф. (ограничимся комбинаторами,
то есть замкнутыми термами, или выражениями без свободных
переменных) можно получить некоторую начальную часть
префиксного кода, не анализируя, в общем случае, еще не
вычисленные подвыражения.
\begin{definition}
Под \textit{префиксным кодом} будем понимать код, составленный
для н.~ф. (он должен совпадать для конвертируемых термов),
с грамматикой
\begin{verbatim}
<t> ::= 1 <t> <t> | 0
\end{verbatim}
Причем ${1\, t_L\, t_R}$ означает ориентированное бинарное дерево
без нагрузки в узлах с левой ветвью в виде дерева ${t_L}$
и правой~--- ${t_R}$.
\end{definition}
\begin{theorem}
Для любой г.~н.~ф. можно составить некоторое непустое начало
префиксного кода.
\end{theorem}
\begin{proof}
Любая г.~н.~ф. (головная нормальная форма) имеет вид
(см. \cite{monograph})
${\lambda \overline x. x\, \overline M}$, где
$\overline x$~--- последовательность переменных ($x$~--- одна
из них ввиду замкнутости терма) длиной $m$, а $\overline M$~---
последовательность выражений длиной $n$, причем ${m > 0}$,
а ${n \ge 0}$.
Рассмотрим возможные случаи.
\begin{enumerate}
\item ${m = 1}$, и г.~н.~ф. имеет вид $$
\lambda x. P\, Q = S\, (\lambda x. P)\, (\lambda x. Q) =
((X\, (X\, X))\, (\lambda x. P))\, (\lambda x. Q)),
$$ т.~е. начало кода~--- ${1110100}$.
\item ${n = 0}$, и г.~н.~ф.~--- тривиальная н.~ф., для которой код
можно получить полностью.
\item ${n > 0}$, и г.~н.~ф. становится $$
\lambda \overline x'. \lambda x_n. P\, Q =
\lambda \overline x'. S\, (\lambda x_n. P) (\lambda x_n. Q),
$$ которая, в свою очередь, попадает либо под случай (1), либо
под (3) с уменьшенной длиной последовательности переменных.
\end{enumerate}

Таким образом, начало кода оказывается восстановимым.
\end{proof}

%% file: nequal.tex
\section{Определение $n$-равенства}

При последующей редукции г.~н.~ф., которая по сути является
списком выражений ${\overline M = \overline M'\, M_n}$, г.~н.~ф.
могут принять и выражения ${\overline M'}$ по мере того,
насколько осмысленны выражение и соответствующий ему алгоритм.
Тогда префиксный код для сравнения можно продолжить, и, кстати,
таким образом неожиданно принимает линейный вид приложение
к методу вывода-вывода по схеме $$
\mathrm{Output} = (\lambda i. \mathrm{Program})\, \mathrm{Input}
$$ с помощью бесконечных списков.
Схема будет пояснена позднее, когда мы перейдем к проекту
<<Macro Lambda Calculus>> и машине редукции для него.

В связи с этим введем следующее определение отношения термов.
\begin{definition}
Термы \textit{$n$-равны}, если равны первые $n$ бит
префиксного кода.
\end{definition}
Такое понятие потенциально могло бы быть полезным для сравнения
рекурсивных алгоритмов.
(Однако, правые части аппликационных термов вида ${Y\, P}$ обычно
позволяют проверку на равенство путем получения
${\beta\eta}$-н.~ф.)
Действительно, соответствующие им $\lambda$-термы принципиально
не могут иметь н.~ф., но исходя из осмысленности (разрешимости)
таких термов, они должны иметь г.~н.~ф.
При неизбежном недостатке памяти машины редукции, на определенном
шаге редукции можно заявить о <<равенстве>> алгоритмов.

%% file: ext.tex
\section{Проблема с экстенсиональностью}

Рассмотрим суть проблемы с экстенсиональностью, обозначенной
ранее, и покажем, какие полезные свойства можно было бы извлечь
из ее решения.

Заметим, что даже если мы будем откладывать составление
префиксного кода выражения ${\lambda x. P\, Q}$, пока не будет
получена г.~н.~ф. $Q$, мы немедленно обнаружим контрпример:
${\lambda x y. x\, (\lambda z. y\, z)}$.
Модифицировать эту идею так, чтобы она действительно решала
упомянутую проблему не представляется возможным, так как в общем
случае составленный код для г.~н.~ф. может из-за последующей
$\eta$-редукции измениться до самого первого бита.

Решению данной проблемы уделяется столько внимания по следующей
причине.
Получение кодов экстенсиональной н.~ф. во время вычисления
позволило бы реализовать полноценный язык программирования без
дополнения чистого бестипового $\lambda$-исчисления какими-либо
$\delta$-функциями или операциями с побочными действиями.
Под полноценностью здесь понимается ввод-вывод данных в виде
нумералов, списков и т.~п.
(Это не отменяет, конечно, эзотеричности языка программирования
в общем случае.)
Человекочитаемого формата можно добиться с помощью сопоставления
формы результирующего кода с образцами.
Последнее является просто частным случаем рассмотренных ранее
словарей нормальных форм, для которых и требуется сериализация
с обязательной экстенсиональностью, так как без последней в общем
случае сопоставление с образцами невозможно без замены
$\eta$-редукции эквивалентным преобразованием.

Сериализация прямо во время вычисления, в свою очередь, обладает
тем полезным свойством, что можно вывести готовую часть списка до
получения его полного вида, или, к примеру, сообщить, больше
какого натурального числа будет еще не готовый ответ к задаче.

%% file: unlimnf.tex
\section{Неограниченная нормализация}

Если разрешить проблему с экстенсиональностью в алгоритме
построения префиксного кода по ходу вычисления выражения, то можно
ввести следующие два определения.
(Вопрос о том, возможно ли решение обозначенной проблемы,
остается открытым.)
\begin{definition}
Назовем \textit{неограниченно нормализуемыми} термами те,
для которых можно составить префиксный код либо полный конечный
(в случае существования н.~ф.), либо бесконечный.
\end{definition}
Данное свойство сильнее существования г.~н.~ф.,
но слабее существования н.~ф.
Примером обладающего таким свойством терма,
может служить выражение, описывающее список всех замкнутых н.~ф.
(это множество рекурсивно перечислимо, но не рекурсивно)
на основе комбинатора $E$, который возвращает $n$-ный комбинатор
(на самом деле, существует целый класс таких термов,
см. \cite{monograph}): $$
Y\, (\lambda f n.[E\, n, f\, (S^+\, n)])\, \lceil 0 \rceil,
$$ где ${Y = \Theta = A\, A}$, причем
${A = \lambda x y. y\, (x\, x\, y)}$
(вариант комбинатора неподвижной точки по Тьюрингу);
${[M, N] = \lambda z. z\, M\, N}$ (спаривание);
${\lceil 0 \rceil = I = \lambda x. x}$,
и ${S^+ = \lambda x. [F, x]}$, причем ${F = \lambda x y. y}$
(ложь).

(Напомним, что нуль ${\lceil 0 \rceil}$, функция следования
${S^+}$ и тест на нуль ${\mathrm{Zero} = \lambda x. x\, T}$,
где истина ${T = \lambda xy.x}$, составляют
\textit{стандартную цифровую систему}, а функция предшествования
${P^- = \lambda x. x\, F}$ делает данную цифровую систему
\textit{адекватной}.)

Данное определение было бы полезно как формализация
<<достаточно осмысленных>> выражений.
Дополнительно к нему введем более сильное свойство,
связав только что введенное с количеством редукций,
требуемым для получения очередного одного бита префиксного кода.
\begin{definition}
Определим $\lambda$-термы \textit{неограниченно нормализуемыми
в реальном времени}, когда существует некоторое конечное число
редукций, необходимых для получения любого очередного бита
префиксного кода.
\end{definition}
(В случае с планируемой системой MLC каждый шаг редукции сам
выполняется за реальное время, поэтому вариант определения с
вычислением на основе такой системы становится относящимся к
фактическому <<real time>>.)
Предыдущий пример не обладает вторым свойством ввиду растущей
длины элементов списка, поэтому иллюстрацией нормализуемого
неограниченно в реальном времени терма нам послужит другой его
вариант (бесконечный список с чередующимися истиной и ложью): $$
Y\, (\lambda f. [T, F, f]),
$$ где ${[M_1, \dots , M_n] = [M_1, [M_2, \dots , M_n]]}$,
а ${[M] = M}$.

Остается открытым вопрос о том, является ли Тьюринг-полной
система, ограниченная нормализуемыми (в реальном времени)
термами.

%% file: btio.tex
\section{Система ввода-вывода на бинарных деревьях}

Продолжим использовать обозначения, определения и утверждения,
сделанные ранее и рассмотрим более детально предложенную систему
ввода-вывода по схеме $$
\mathrm{Output} = (\lambda i. \mathrm{Program})\, \mathrm{Input}.
$$
Для этого воспользуемся введенной сериализацией
$\lambda$-выражений и тривиальной десериализацией: она получается
подстановкой $X$, составляющего одноточечный базис комбинаторной
логики, вместо листьев ориентированного бинарного дерева
без нагрузки в узлах, кодируемого префиксным кодом, и заменой
узлов на аппликации левой их ветви к правой.
А именно, отныне будем считать, что ${\mathrm{Input}}$
первоначально представляет собой сериализованный (возможно,
бесконечный) список, подлежащий десериализации по мере
необходимости для редукций, требуемых, в свою очередь, для
получения префиксного кода ${\mathrm{Output}}$.

Тогда существует система кодирования для текстов
(последовательностей символов из некоторого конечного набора)
в виде определенных нами префиксных кодов, такая, что любая
программа в смысле алгоритма преобразования одного текста в
другой может быть закодирована термом
${\mathrm{Program} \in \Lambda^0(i)}$, включая те программы,
которые работают с бесконечными текстами (как, например,
\texttt{cat(1)} и \texttt{grep(1)} в \cite{posix}), причем
программы оказываются всюду определенными, если выражения
${\mathrm{Program}}$ и ${\mathrm{Input}}$ неограниченно
нормализуемы, так как из этого следует также неограниченная
нормализация ${\mathrm{Output}}$.
\begin{theorem}
Если $\lambda$-термы $M$ и $N$ неограниченно нормализуемы,
то неограниченно нормализуемо и выражение ${M\, N}$.
\end{theorem}
\begin{proof}
Действительно, если $M$ и $N$~--- г. н. ф.,
то и $M\, N$ тоже разрешим, а так как г. н. ф. поочередно принимают
все элементы последовательности списка подвыражений в г. н. ф.,
то префиксный код не обрывается без завершающего нуля.
\end{proof}

Системы реального времени, в свою очередь, соответствуют термам
${\mathrm{Program}}$, неограниченно нормализуемым в реальном времени,
если кодирование текстов выбрано так, что ${\mathrm{Input}}$ и,
следовательно, ${\mathrm{Output}}$, обладают тем же свойством.

%% file: uniform.tex
\section{Однородная память как особый вид графов}

Однородность памяти в описании машины редукции
<<The Heap Lambda Machine>> \cite{heap}
может быть обобщена на особый вид систем с памятью устроенной
графами определенной структуры.
Ниже будет построено обобщение используемой памяти, затем будут
введены свойства систем на ее основе, и, в частности, показано,
что <<The Heap Lambda Machine>> является частным случаем таких
систем, а планируемая машина редукции в проекте MLC обладает
более сильными свойствами.
\begin{definition}
Будем называть \textit{однородной памятью} ориентированный
направленный граф (то есть такой граф, дуги, исходящие из любого
узла которого составляют упорядоченное множество),
с $N$ узлами, каждый из которых имеет ровно ${N_C \ge 1}$
исходящих из него дуг.
\end{definition}
Поясним выбор названия для конструкции и имя индекса в ${N_C}$:
естественным в компьютерном мире примером определенной
конструкции может служить память, состоящая из $N$ одинаковых
блоков со структурой в виде массива из ${N_C}$ ячеек
${c_0 \dots c_{N_C}}$ (англ. cells), в которых содержатся ссылки
на другие блоки в рамках этой памяти.
Но несмотря на то, что в компьютерах блоки можно было бы
использовать также для хранения и обработки содержимого ячеек
как натуральных чисел, а не указателей, и определять также
строки, числа с плавающей точкой и~т.~п. при должном ALU и FPU,
нам будут интересны лишь системы без подобных механизмов
в процессоре или процессорах; также мы не будем пока
рассматривать ввод-вывод.

Теперь рассмотрим возможные свойства систем на основе однородной
памяти.
\begin{definition}
Более точно, под \textit{системами} мы будем подразумевать машины
с состоянием в виде однородной памяти как вычислимые функции
на однородной памяти с примитивными операциями, ограниченными
сравнением двух нод и изменением дуг, исходящих из данной ноды.
\end{definition}

%% file: property.tex
\subsection{Свойства систем на однородной памяти}

\begin{definition}
Системой \textit{без потери памяти} мы будем называть
такие системы, состоянием которых являются только связанные
графы, тогда и только тогда, когда существует алгоритм, который
по данному некоторому выделенному узлу ${\mathrm{NULL}}$ и
ссылкам из него выстраивает все узлы в список через первую ячейку
${c_0}$, последний элемент которого имеет
${c_0 = \mathrm{NULL}}$.
\end{definition}
\begin{definition}
\textit{Тьюринг-полнотой} таких систем назовем Тьюринг-полноту
этих же систем, положив ${N \rightarrow \infty}$.
\end{definition}
Можно сказать, что Тьюринг-полные системы в нашем смысле являются
реализациями Тьюринг-полных систем в общем смысле.
Примером Тьюринг-полной системы без потери памяти и является
упомянутая в начале раздела <<The Heap Lambda Machine>>
с ${N_C = 4}$.
\begin{definition}
Работой \textit{в реальном времени} системы
заявим способность системы переходить из любого состояния к
следующему путем изменения и анализа конечного связанного
подграфа, начинающегося с ${\mathrm{NULL}}$, то есть с
количеством узлов, ограниченным некоторым конечным числом.
\end{definition}
Комбинацией всех трех введенных свойств и должна обладать
планируемая машина редукции в проекте MLC; пока выбрано
${N_C = 4}$, как и в случае с <<The Heap Lambda Machine>>, но
интересным оказывается нахождение минимальных значений ${N_C}$
для Тьюринг-полной системы без потери памяти как с работой
в реальном времени, так и без нее.

%% file: strategy.tex
\section{Нормализация в дистрибутивном $\lambda$-исчислении}

Нормальная стратегия не является нормализующей для
дистрибутивного $\lambda$-исчисления.
Тем не менее, существует по крайней мере одна эффективная
(рекурсивная) нормализующая стратегия для такой редукции~---
это так называемая внутренняя спиновая стратегия, которая
отличается от нормальной лишь в выборе внутренней абстракции
${(\lambda x. M)\, N}$ после редукции (3) (см. первую часть
обоснования машины редукции MLC), даже если внешняя
абстракция ${\lambda y. (\lambda x. M)\, N}$ участвует в новом
$\beta$-редексе.
Машина будет реализовывать некоторую вариацию этой стратегии,
не слабее, чем теоретический прототип.

В приложении будет приведена стратегия с доказательством ее
нормализующего свойства.
Стратегия изначально была предложена автором в неформальном виде
и в виде компьютерной программы, а затем формализована
Vincent van Oostrom в личной переписке.
Доказательство закрывает прежде открытый вопрос в \cite{ustica}.

%% file: mlc.tex
\section{Проект <<Macro Lambda Calculus>>}

\begin{verbatim}
<text> ::= <term> | <assign> <text>
<assign> ::= <ID> '=' <term> ';'
<term> ::= <appl> | <abstr>
<abstr> ::= <ID> ':' <term> | <ID> ',' <abstr>
<appl> ::= <atom> | <appl> <atom>
<atom> ::= <ID> | '(' <term> ')'
\end{verbatim}

Грамматика, с которой начался данный раздел, описывает
синтаксис языка программирования в проекте
<<Macro Lambda Calculus>>.
Легко заметить, что язык представляет собой простую форму
чистого бестипового $\lambda$-исчисления с макроопределениями.
Для данного языка планируется разработать интерпретатор в виде
машины редукции определенного вида, который будет пояснен в
последующих разделах.

В рамках MLC рассматривается ввод-вывод по схеме $$
\mathrm{Output} = (\lambda i. \mathrm{Program})\, \mathrm{Input},
$$ подробно разобранной выше.
Предполагается, что во время вычисления при проверке
на присутствие редекса в нодах, где ${\mathrm{Input}}$ или
какое-либо его подвыражение стоит в левой части аппликации,
оставляется запрос на чтение, редекс откладывается, после чего
вычисление выражения продолжается, а позже, когда левая часть
аппликации будет достроена, то есть считана, список редексов
обновляется так, чтобы следующим редексом оказался именно он.

%% file: sugar.tex
\subsection{<<Синтаксический сахар>>}

Покажем пример использования введенного языка с комментариями
(строки, начинающиеся с <<\verb'---'>> игнорируются), в котором 
переопределим логические значения и операции, списки, комбинатор
неподвижной точки, нумералы, простейшие арифметические операции,
а затем представим в языке выражение для ${4!}$ (факториал $4$).
\begin{verbatim}
--- Identity, the simplest combinator
I = x: x;

--- Select either first or second
TRUE = first, second: first;
FALSE = first, second: second;

--- If cond1 then cond2 else FALSE
AND = cond1, cond2: cond1 cond2 FALSE;
--- If cond1 then TRUE else cond2
OR = cond1, cond2: cond1 TRUE cond2;
--- Swap arguments for TRUE to become FALSE and vice versa
NOT = cond: (first, second: cond second first);

--- Pair of head and tail
LIST = head, tail: (x: x head tail);
--- List operations are postfix
LISTOP = op: (list: list op);
--- Select first part of pair
HEAD = LISTOP TRUE;
--- Select second part of pair
TAIL = LISTOP FALSE;
--- Ignore both parts of pair and return FALSE
NULL = LISTOP (first, second: FALSE);
--- Pseudo-list that ignores list operation and returns TRUE
NIL = x: TRUE;

--- Lists of identity terms ended with NIL
N0 = NIL;
N1 = LIST I N0;
N2 = LIST I N1;
N3 = LIST I N2;
N4 = LIST I N3;
--- Increment list length
SUCC = LIST I;
--- Decrement list length
PRED = TAIL;
--- If N0 then TRUE else FALSE
ZERO = NULL;

--- Fixed point combinator by Turing
A = self, func: func (self self func);
Y = A A;

--- Addition defined recursively for any numeral system
PLUSR = self, m, n: (ZERO m) n (SUCC [self {PRED m} n]);
PLUS = Y PLUSR;
--- Multiplication defined recursively for any numeral system
MULTR = self, m, n: (ZERO m) N0 (PLUS m [self {PRED m} n]);
MULT = Y MULTR;
--- Traditional factorial function example
FACTR = self, n: (ZERO n) N1 (MULT n [self {PRED n}]);
FACT = Y FACTR;

--- Evaluate factorial of four
FACT N4
\end{verbatim}

%% file: overview.tex
\section{Описание машины MLC}

В проекте <<Macro Lambda Calculus>> машина представляет собой
однородную память, адресуемую линейно, и процессор без
арифметико-логического устройства, работающий по тактам, то есть
время каждого шага его работы ограничено некоторой константой~---
ниже будет использоваться формулировка <<в реальном времени>>,
подразумевающая, что действие выполняется за время ${O(1)}$.
В памяти находится бестиповое $\lambda$-выражение с так
называемым <<ленивым>> связыванием подвыражений и кольцевые
двусвязные списки редексов и свободных нод.

Каждый адресуемый блок памяти имеет размер четырех адресов,
которые используются по-разному, в зависимости от контекста.
Каждая нода выражения снабжается собственным двусвязным списком
использований соответствующего подвыражения ввиду <<ленивого>>
связывания.

Перед вычислением выражения, то есть до начала работы машины,
требуется подготовить начальное состояние списка редексов.
В него входят все $\beta$- и $\eta$-редексы в выражении в
порядке, диктуемом нормальной стратегией, а также следующий ряд
подвыражений.

Пусть вектор ${\overline M = M_1, \dots , M_n}$, и $R$~---
$\beta$-редекс, причем ${R\, \overline M}$~--- подвыражение
вычисляемого системой выражения.
Тогда в список редексов добавляются также каждое из подвыражений
${R\, M_1}, {R\, M_1\, M_2, \dots , R\, \overline M}$ в таком же
порядке сразу после редекса $R$.

После загрузки выражения в память машины и подготовки начального
состояния списка редексов машина может начать свою работу.
Каждый шаг работы производит анализ первого элемента списка
редексов.
Если элемент ссылается на подвыражение, которое действительно
является редексом, производится один шаг редукции по правилам
дистрибутивного $\lambda$-исчисления.
Затем список редексов обновляется за время, ограниченное
константой, так, чтобы обеспечить нормализующую стратегию.

Реализация правил дистрибутивного $\lambda$-исчисления требует
удаления некоторых подвыражений и выделения новых нод.
Так как требуется реализация в реальном времени, то для удаления
и выделения нод был выбран своеобразный <<сборщик мусора>>.

По сути, удаление подвыражения приводит лишь к удалению
соответствующего элемента списка использований без прохода вниз
по дереву этого подвыражения.
После удаления ноды, она заносится в список свободных нод.
При этом дерево соответствующего подвыражения в общем случае все
еще сохраняется.
Когда приходит время выделять ноду, ее дочерние элементы проходят
ту же процедуру удаления.
Таким образом, удаление подвыражений оказывается реализуемым в
реальном времени без потерь.

%% file: plan.tex
\subsection{План обоснования}

Утверждается, что машина построена корректно и действительно
находит существующие ${\beta\eta}$-н.~ф. при достаточной памяти.
Следовательно, нам требуется обосновать утверждение о том,
что механизм работы машины построен корректно, обеспечивается
работа в реальном времени всех шагов и машина действительно
обладает нормализующими свойствами при экстенсиональности
без потери блоков памяти.
Обоснование будет состоять из нескольких частей.

\begin{enumerate}
\item Предлагаемая реализация правил дистрибутивного
$\lambda$-исчисления, которые заменяют $\beta$-редукцию.
\item Метод обнаружения новых $\eta$-редексов.
\item Доказательство достаточности начального состояния списка
редексов для экстенсиональной нормализации вычисляемого
выражения (при достаточной памяти и существовании
${\beta\eta}$-н.~ф.).
\end{enumerate}

%% file: first.tex
\subsection{Первая часть}

Рассмотрим все четыре правила редукции дистрибутивного
$\lambda$-исчисления.
Напомним, что правила покрывают четыре возможных формы тела
функции в $\beta$-редексе.

\begin{enumerate}
\item ${(\lambda x. x)\, M = M}$ (тождество).
\item ${(\lambda x. y)\, M = y}$ (константа).
\item ${(\lambda x y. M)\, N = \lambda y. (\lambda x. M)\, N}$
(абстракция).
\item ${(\lambda x. M_1\, M_2)\, N =
(\lambda x. M_1)\, N\, ((\lambda x. M_2)\, N)}$ (аппликация).
\end{enumerate}

(Дублирование переменной может привести к потере $\eta$-редексов
и невозможности их нахождения по количеству использований
переменной~--- требует отдельного рассмотрения; возможно,
корректность обеспечивается тем, что переменная $x$ уже не будет
участвовать в $\eta$-редексах, которые бы оставались после
достижения $\beta$-н.~ф.)

Итак, перейдем к описанию реализации каждого из правил.
Возможные новые $\beta$-редексы, появляющиеся в контекстах,
использующих редуцируемую ноду, будут рассмотрены позднее.
Следует иметь в виду, что замена ноды редекса на другую ноду
сопровождается созданием дополнительного элемента списка ее
использований.
В каждом случае редекс как подвыражение удаляется.
Под <<созданием>> ноды или (соответствующего) подвыражения
подразумевается выделение не больше некоторого постоянного числа
блоков памяти с увеличением количества использований участвующих
подвыражений.

\begin{enumerate}
\item Нода редекса заменяется на ноду $M$.
Текущий элемент списка редексов удаляется.
Новых редексов в редуцируемом подвыражении не создается.
\item Нода редекса заменяется на ноду $y$.
Текущий элемент списка редексов удаляется.
Новых редексов в редуцируемом подвыражении не создается.
\item Создаются ноды ${\lambda x. M}$, ${(\lambda x. M)\, N}$
и ${\lambda y. (\lambda x. M)\, N}$, и нода редекса заменяется
на последнюю.
Текущий элемент списка редексов заменяется на
${(\lambda x. M)\, N}$, что диктуется внутренней спиновой
стратегией.
Больше новых редексов в получившемся подвыражении не создается.
\item Создаются ноды ${\lambda x. M_1}$, ${\lambda x. M_2}$,
${(\lambda x. M_1)\, N}$, ${(\lambda x. M_2)\, N}$ и
${(\lambda x. M_1)\, N\, ((\lambda x. M_2)\, N)}$, и нода редекса
заменяется на последнюю.
Перед текущим элементом списка редексов вставляется
${(\lambda x. M_1)\, N}$ (возможно, стратегия нарушится, если
${M_1}$ является аппликацией), а в конец списка (что возможно
из-за цикличности списка) редексов~--- ${(\lambda x. M_2)\, N}$
(недостает доказательства того, что последнее сохраняет
нормализующее свойство получившейся стратегии).
Больше новых редексов в получившемся подвыражении не создается.
\end{enumerate}

%% file: second.tex
\subsection{Вторая часть}

Заметим, что $\eta$-редексы могут порождать новые $\eta$-редексы:
$$
\lambda x. y\, (\lambda z. x\, z) = \lambda x. y\, x.
$$
В свою очередь, $\beta$-редексы также могут порождать
$\eta$-редексы: $$
\lambda x. y\, (I\, x) = \lambda x. y\, x.
$$
Обратное последнему утверждение, однако, неверно.
Покажем это.
\begin{theorem}
Контракция $\eta$-редексов не порождает новых $\beta$-редексов.
\end{theorem}
\begin{proof}
Пусть ${H = \lambda x. M\, x}$ есть $\eta$-редекс, то есть $M$
не содержит свободных вхождений переменной $x$.
Возьмем произвольный контекст ${C[\phantom M]}$ и рассмотрим три
исчерпывающих взаимоисключающих варианта (${C'[\phantom M]}$~---
некоторый контекст, и $N$~--- произвольный терм).
\begin{enumerate}
\item ${C[\phantom M] = C[[\phantom M]\, N]}$.
\item ${C[\phantom M] = C'[N\, [\phantom M]]}$.
\item ${C[\phantom M] = C'[\lambda x. [\phantom M]]}$.
\end{enumerate}
Очевидно, что для случаев (2) и (3) $\eta$-редукция $H$ к $M$ в
выражении ${C[H] = C[M]}$ не приводит к появлению новых редексов.
В случае (1) ${M\, N}$ может оказаться $\beta$-редексом, однако
выражение ${H\, N}$ являлось $\beta$-редексом еще до
$\eta$-редукции $H$ к $M$.
\end{proof}

Из этого факта следует, что безопасно откладывать $\eta$-редукцию
сколь угодно долго без угрозы нормализующему свойству.
Таким образом, оказывается приемлемым решение обнаруживать новые
$\eta$-редексы при процедуре удаления нод, отвечающих
использованиям переменной, то есть в общем случае к концу работы
<<сборщика мусора>>.
(Подразумевается, что в случае завершения
обработки списка редексов, <<сборщик мусора>> начинает выделять
свободные блоки памяти, тем самым выполняя требуемую проверку
переменных, очищать их, и включать обратно в список свободных
нод, проходя по нему полностью, пока не останется никаких
деревьев подвыражения, подлежащих удалению.)

Проверка на редекс показывает положительный результат тогда и
только тогда, когда удаляется использование переменной, и
остается еще два, причем оба они укладываются в форму
$\eta$-редекса.
Это можно определить в реальном времени благодаря ссылкам на
родительские ноды в элементах списков использований.

Любые переменные, которые могут участвовать в новых
$\eta$-редексах, оказываются в конечном итоге в списке свободных
нод, а значит и на проверке <<сборщиком мусора>>, так как после
каждой редукции удаляется редекс, и он является единственной
изменяемой частью вычисляемого выражения.

%% file: third.tex
\subsection{Третья часть}

Напомним, какие ноды заносятся в список редексов до начала работы
системы.
Это все $\eta$-редексы в произвольном порядке, все
$\beta$-редексы в порядке, диктуемом нормальной стратегией, но
сразу после каждого $\beta$-редекса $R$ заносятся также ноды $$
R\, M_1, R\, M_1\, M_2, \dots , R\, \overline M,
$$ где $\overline M$~--- вектор наибольшей длины $n$ (возможно,
нулевой), который удовлетворяет структуре контекста
${C[\phantom M]}$ редекса $R$.
Подразумевается, что для $R$ и всех элементов вектора
$\overline M$ ноды, подлежащие занесению в список редексов, были
или будут обработаны отдельно от цепочки аппликаций
${R\, \overline M}$.
Покажем, что начальное состояние списка редексов, подготовленное
таким образом, достаточно для сохранения нормализующего свойства
в реальном времени машиной с описанными ранее структурой и
механизмами работы.
\begin{theorem}
Начальное состояние и механизм работы машины MLC достаточны для
нормализующего свойства.
\end{theorem}
\begin{proof}
Так как вектор $\overline M$ имеет наибольшую длину, то контекст
${C[\phantom M]}$ может иметь лишь одну из следующих форм, где
$L$ есть некоторый терм, а ${C'[\phantom M]}$~--- более внешний
контекст.
\begin{enumerate}
\item ${C[\phantom M] = C'[\lambda x. [\phantom M]]}$.
\item ${C[\phantom M] = C'[L\, [\phantom M]]}$.
\item ${C[\phantom M] = [\phantom M]}$.
\end{enumerate}
При этом форма ${C'[[\phantom M]\, L]}$ невозможна, так как длина
вектора $M$ выбрана максимальной, а случай (3) тривиален.

Следует также отметить, что остальные ноды, кроме цепочки
аппликаций ${R\, \overline M}$, в контексте ${C'[\phantom M]}$
также обрабатываются отдельно.
Этого достаточно потому, что редексом $R$ могут порождаться новые
$\beta$-редексы лишь в подвыражении ${R\, \overline M}$ ввиду
ограничения области создания новых $\beta$-редексов телом
абстракции в случае (1) и правой частью аппликации в (2).

При этом вышеописанный механизм работы самих шагов редукции
контролирует более внутренние возможные редексы и на одном с $R$
уровне.
Значит, сейчас требовалось действительно рассмотреть лишь те
редексы, что создаются в родительских нодах.
\end{proof}

%% file: end.tex
\section{Заключение}

Итак, мы обозначили определенный класс систем~--- машин редукции
графов с однородной памятью и ввели способ сериализации
результатов не после завершения вычисления, которого может и не
быть вовсе, а во время выполнения редукции.
Также мы определили классы термов, для которых процесс
сериализации обладает полезными свойствами.

По мере развития поднятой в данной статье темы, остались
открытыми некоторые интересные вопросы: например, о
Тьюринг-полноте введенных классов термов и даже об
эквивалентности неограниченно нормализуемых и неограниченно
нормализуемых в реальном времени термов;
не менее волнующим автора вопросом является и минимальные
количества ячеек в блоках однородной памяти для выполнения
различных комбинаций из трех рассмотренных свойств систем
на ее основе.

На данном этапе, тем не менее, приоритет отдается реализации
машины MLC как интерпретатора языка программирования, обладающей
всеми тремя введенными свойствами, в рамках стандартной
универсальной операционной системы POSIX \cite{posix} с помощью
таких средств, как \texttt{c99(1)}, \texttt{lex(1)},
\texttt{yacc(1)} и \texttt{make(1)}.

%% file: appendix.tex
\appendix
\section{Внутренняя спиновая стратегия}

Мы положительно закрываем вопрос A.1.6 в \cite{ustica} о том,
существует ли рекурсивная нормализующая одношаговая стратегия
для <<Micro Lambda Calculus>>, где последнее обозначает
реализацию $\lambda$-исчисления по R\'ev\'esz \cite{revesz},
разбивая (англ. distribute) $\beta$-редукцию на четыре
исчерпывающих простейших правила в зависимости от формы тела $M$
абстракции в $\beta$-редексе ${(\lambda x. M) N}$.
Но сначала мы снабдим редукцию <<Micro Lambda Calculus>> более
формальным наименованием и дадим ей строгое определение\footnote{
Vincent van Oostrom предложил личным сообщением \cite{private}
данные построения в несколько другом виде~--- автор позволил себе
видоизменить обозначения и выражает надежду на то, что ему
удалось упростить доказательства двух важных утверждений.}.
\begin{definition}
\textit{Дистрибутивная редукция} определяется как $$
\beta_d = \beta_d^i \cup \beta_d^c \cup \beta_d^l \cup \beta_d^a,
$$ где
\begin{eqnarray*}
\beta_d^i &=& \bigl\{((\lambda x. x)\, M, M)
\bigm| M \in \Lambda\bigr\}, \\
\beta_d^c &=& \bigl\{((\lambda x. y)\, M, y)
\bigm| M \in \Lambda\bigr\}, \\
\beta_d^l &=& \bigl\{((\lambda x. \lambda y. M)\, N,
\lambda y. \lambda x. M\, N)
\bigm| M, N \in \Lambda\bigr\}, \\
\beta_d^a &=& \bigl\{((\lambda x. M\, N)\, P,
(\lambda x. M)\, P\, ((\lambda x. N)\, P))
\bigm| M, N, P \in \Lambda\bigr\}.
\end{eqnarray*}
Дополнительно обозначим следующие отношения:
\begin{eqnarray*}
M \rightarrow_d N &\Leftrightarrow& (M, N) \in \beta_d, \\
M \rightarrow_i N &\Leftrightarrow& (M, N) \in \beta_d^i, \\
M \rightarrow_c N &\Leftrightarrow& (M, N) \in \beta_d^c, \\
M \rightarrow_l N &\Leftrightarrow& (M, N) \in \beta_d^l, \\
M \rightarrow_a N &\Leftrightarrow& (M, N) \in \beta_d^a.
\end{eqnarray*}
\end{definition}
\begin{theorem}
Любой ${\beta_d}$-редекс есть $\beta$-редекс, и наоборот.
\end{theorem}
\begin{proof}
Утверждение непосредственно следует из определения ${\beta_d}$.
\end{proof}

Нормализующее свойство стратегии, которая и будет закрывать
прежде открытый вопрос, опирается, с одной стороны, на
нормализующее свойство спиновой редукции для обычного
$\lambda$-исчисления, а с другой~--- на завершимость элементарных
правил, как в системе ${\lambda x}$~--- варианте
$\lambda$-исчисления с <<explicit substitution>>.
Определим саму стратегию, для которой мы будем доказывать
нормализующее свойство.
\begin{definition}
\textit{Внутренняя спиновая стратегия} всегда отдает приоритет
самому внутреннему редексу среди спиновых редексов (см. определение
4.7 в \cite{needred}).
\end{definition}

\input correct
\input def
\input proof

%% file: correct.tex
\subsection{Корректность дистрибутивной редукции}

Покажем, что дистрибутивная редукция сохраняет конвертируемость.
\begin{theorem}
${M \rightarrow_d N \Rightarrow M =_\beta N}$.
\end{theorem}
\begin{proof}
Рассмотрим каждое из четырех подмножеств ${\beta_d}$.
\begin{enumerate}
\item Если ${M \rightarrow_i N}$, то для некоторого $P$ $$
M \equiv (\lambda x. x)\, P \land N \equiv P,
$$ но тогда $$
(\lambda x. x)\, P \rightarrow_\beta x[x := P] \equiv P.
$$
\item Если ${M \rightarrow_c N}$, то для некоторого $P$ $$
M \equiv (\lambda x. y)\, P \land N \equiv y,
$$ но тогда $$
(\lambda x. y)\, P \rightarrow_\beta y[x := P] \equiv y.
$$
\item Если ${M \rightarrow_l N}$, то для некоторых $P$ и $Q$ $$
M \equiv (\lambda x. \lambda y. P)\, Q \land
N \equiv \lambda y. (\lambda x. P)\, Q,
$$ но тогда $$
(\lambda x. \lambda y. P)\, Q \rightarrow_\beta
(\lambda y. P)[x := Q] \equiv \lambda y. P[x := Q]
\leftarrow_\beta \lambda y. (\lambda x. P)\, Q.
$$
\item Если ${M \rightarrow_a N}$, то для некоторых $P$, $Q$ и $R$
$$
M \equiv (\lambda x. P\, Q)\, R \land
N \equiv (\lambda x. P)\, R\, ((\lambda x. Q)\, R),
$$ но тогда $$
(\lambda x. P\, Q)\, R \rightarrow_\beta
(P\, Q)[x := R] \equiv P[x := R]\, Q[x := R]
\leftarrow_\beta (\lambda x. P)\, R\, ((\lambda x. Q)\, R),
$$
\end{enumerate}
Так как мы рассмотрели ${\beta_d^i}$, ${\beta_d^c}$,
${\beta_d^l}$ и ${\beta_d^a}$, то утверждение верно и для
${\beta_d}$.
\end{proof}

%% file: def.tex
\subsection{Необходимые определения}

\begin{definition}
${M^\bullet}$~--- это терм, полученный контракцией всех
$\beta$-редексов $M$.
Операцию $\bullet$ можно рассматривать как
\textit{полное $\beta$-развитие} терма $M$.
\end{definition}
\begin{definition}
Шаг называется \textit{деструктивным}, если он производит
контракцию редексов вида
${((\lambda x. (\lambda y. P)\, Q))\, R}$.
Остальные шаги назовем \textit{недеструктивными}.
\end{definition}
Наша стратегия основана на наблюдении, что дистрибутивная
редукция сохраняется при проектировании каждого терма $M$
на его ${M^\bullet}$, коль скоро ее шаги не являются
деструктивными.
Недеструктивные шаги будут спроектированы с помощью
операции полного $\beta$-развития на $\beta$-редукционные
цепочки.

Однако, вместо того, чтобы доказывать настолько общее
утверждение, мы заметим, что шаги, диктуемые внутренней
спиновой стратегией (для краткости такие шаги мы будем называть
внутренними спиновыми шагами), недеструктивны благодаря тому, что
они отдают приоритет редукции тела абстракции в выражениях вида
${((\lambda x. (\lambda y. P)\, Q))\, R}$, а затем покажем, что
любой внутренний спиновый шаг соответствует не больше чем одному
шагу $\beta$-редукции.
Более того, в случае, когда внутренний спиновый шаг дистрибутивной
редукции проецируется с помощью $\bullet$ в пустой шаг,
образ не порождал новых $\beta$-редексов, а значит,
такой шаг существует лишь в дистрибутивной редукции.
Последнее может быть представлено формально проецированием такого
шага на $x$-шаги в системе ${\lambda x}$ ($\lambda$-исчисление
с <<explicit substitutions>>, см. \cite{bloo}) с помощью
следующей операции.
\begin{definition}
${M^\diamond}$~--- это терм, полученный заменой каждого
$\beta$-редекса ${(\lambda x. P)\, Q}$ на
${P \langle x := Q \rangle}$ в терме $M$.
Операцию $\diamond$ можно рассматривать как
\textit{эксплификацию} $M$.
\end{definition}

%% file: proof.tex
\subsection{Доказательство нормализующего свойства}

\begin{theorem}
Если ${M \rightarrow_d N}$, то либо шаг
${M^\bullet \rightarrow_\beta N^\bullet}$ выполняет контракцию
спинового редекса, либо
${M^\bullet \equiv N^\bullet}$ и
${M^\diamond \rightarrow_x N^\diamond}$.
\end{theorem}
\begin{proof}
Рассмотрим возможные случаи.
\begin{enumerate}
\item Если внутренний спиновый шаг ${M \rightarrow N}$
соответствует ${M \rightarrow_d N}$, то
${M^\bullet \equiv N^\bullet}$ и
${M^\diamond \rightarrow_x N^\diamond}$ следуют из утверждения
о корректности дистрибутивной редукции.
\input cases
\item Если внутренний спиновый шаг
${\lambda x. M \rightarrow \lambda x. N}$ соответствует
${M \rightarrow_d N}$, то опираясь на доказанность
утверждения для (1), мы немедленно получаем, что либо $$
(\lambda x. M)^\bullet \equiv \lambda x. M^\bullet
\rightarrow_\beta
\lambda x. N^\bullet \equiv (\lambda x. N)^\bullet
$$ является спиновым шагом, либо $$
(\lambda x. M)^\bullet \equiv \lambda x. M^\bullet \equiv
\lambda x. N^\bullet \equiv (\lambda x. N)^\bullet
$$ и $$
(\lambda x. M)^\diamond \equiv \lambda x. M^\diamond
\rightarrow_x
\lambda x. N^\diamond \equiv (\lambda x. N)^\diamond.
$$
\end{enumerate}
Таким образом мы исчерпали все возможные случаи.
\end{proof}
\begin{theorem}
Внутренняя спиновая стратегия является нормализующей.
\end{theorem}
\begin{proof}
На основании предыдущего утверждения, бесконечная дистрибутивная
редукционная цепочка из некоторого терма $M$, имеющего н.~ф.
$\hat M$, породила бы бесконечную спиновую $\beta$-редукционную
цепочку из ${M^\bullet}$, если бы начиная с некоторой формы $N$ в
дистрибутивной цепочке все дальнейшие формы не были бы отображены
на ${N^\bullet}$.
Но тогда снова в силу предыдущего утверждения бесконечная
дистрибутивная цепочка из $N$ породила бы бесконечную
$x$-редукционную цепочку из ${N^\diamond}$.

Бесконечные спиновые $\beta$-редукционные цепочки из
${N^\bullet}$ невозможны, так как $M$ и ${M^\bullet}$ являются
$\beta$-конвертируемыми, таким образом имея одну и ту же
$\beta$-н.~ф. $\hat M$, а спиновые стратегии, в свою очередь,
являются нормализующими \cite{needred}.
В свою очередь, бесконечные $x$-редукционные цепочки невозможны,
так как $x$-редукция (правила подстановки) всегда завершаются в
системе ${\lambda x}$ \cite{bloo}.
\end{proof}
Суть нашей стратегии~--- избежать разрушения редексов:
внутренняя спиновая стратегия запрещает дистрибутивную редукцию
внешнего редекса в ${(\lambda x. (\lambda y. P)\, Q)\, R}$,
блокируя разрушение внутреннего и, следовательно, контрпример
Jan Willem Klop \cite{ustica} к сохранению нормализующего
свойства для дистрибутивной редукции.

%% file: cases.tex
\item Если внутренний спиновый шаг ${M\, P \rightarrow N\, P}$
соответствует ${M \rightarrow_d N}$, то опираясь на доказанность
утверждения для (1), рассмотрим три случая.
\begin{enumerate}
\item ${M\, P}$~--- $\beta$-редекс. \\
Тогда ${M \equiv \lambda x. M'}$ и ${N \equiv \lambda x. N'}$ для
некоторых ${M'}$ и ${N'}$, и ${M' \rightarrow_\beta N'}$,
следовательно, либо $$
(M\, P)^\bullet \equiv M'^\bullet[x := P^\bullet]
\rightarrow_\beta
N'^\bullet[x := P^\bullet] \equiv (N\, P)^\bullet
$$ является спиновым шагом, либо $$
(M\, P)^\bullet \equiv M'^\bullet[x := P^\bullet] \equiv
N'^\bullet[x := P^\bullet] \equiv (N\, P)^\bullet
$$ и $$
(M\, P)^\diamond \equiv
M'^\diamond \langle x := P^\diamond \rangle \rightarrow_x
N'^\diamond \langle x := P^\diamond \rangle \equiv
(N\, P)^\diamond.
$$
\item ${M\, P}$~--- не $\beta$-редекс, но
${N\, P}$~--- $\beta$-редекс. \\
Тогда ${N \equiv \lambda x. N'}$ для некоторого ${N'}$ и
либо ${M \rightarrow_i N}$ и ${M \equiv (\lambda x. x)\, N}$,
либо ${M \rightarrow_l N}$ и для некоторых ${M'}$, ${N''}$
\begin{eqnarray*}
M &\equiv& (\lambda x y. M')\, N''; \\
N' &\equiv& (\lambda x. M')\, N''.
\end{eqnarray*}
Вариант с ${\beta_d^i}$ тривиален, а в случае ${\beta_d^l}$
мы имеем $$
(M\, P)^\bullet \equiv
((\lambda x y. M')\, N'')^\bullet\, P^\bullet \equiv
(\lambda y. M')^\bullet[x := N''^\bullet]\, P^\bullet \equiv
(\lambda y. M'^\bullet[x := N''^\bullet])\, P^\bullet
$$ и $$
(N\, P)^\bullet \equiv ((\lambda y. N')\, P)^\bullet \equiv
(\lambda y. (\lambda x. M')\, N'')\, P^\bullet \equiv
M'^\bullet[x := N''^\bullet][y := P^\bullet]
$$ но тогда ${(M\, P)^\bullet \rightarrow_\beta (N\, P)^\bullet}$
и, следовательно, доказываемое утверждение имеет место, ибо
головной редекс является частным случаем спинового.
\item ${M\, P}$~--- не $\beta$-редекс, и
${N\, P}$~--- тоже не $\beta$-редекс. \\
Тогда ${(N\, P)^\bullet \equiv N^\bullet\, P^\bullet}$ и
${(N\, P)^\diamond \equiv N^\diamond\, P^\diamond}$, из чего
автоматически следует доказываемое утверждение.
\end{enumerate}
\item Если внутренний спиновый шаг ${P\, M \rightarrow P\, N}$
соответствует ${M \rightarrow_d N}$, то опираясь на доказанность
утверждения для (1), заметим, что ${P\, M}$ не может быть
редексом, следовательно, редексом не является и ${P\, N}$.
Но тогда либо $$
(P\, M)^\bullet \equiv P^\bullet\, M^\bullet \rightarrow_\beta
P^\bullet\, N^\bullet \equiv (P\, N)^\bullet
$$ является спиновым шагом, либо $$
(P\, M)^\bullet \equiv P^\bullet\, M^\bullet \equiv
P^\bullet\, N^\bullet \equiv (P\, N)^\bullet
$$ и $$
(P\, M)^\diamond \equiv P^\diamond\, M^\diamond \rightarrow_x
P^\diamond\, N^\diamond \equiv (P\, N)^\diamond.
$$

%% file: paper.bbl
\newpage
\begin{thebibliography}{0}
\bibitem{history} J.~R.~Hindley, F.~Cardone.
History of $\lambda$-calculus and Combinatory Logic.
Handbook of the History of Logic, 5: 723--817, Elsevier, 2009.
\bibitem{sicp} H.~Abelson, G.~J.~Sussman, J.~Sussman.
Structure and interpretation of computer programs. 2nd ed.
New York, NY: McGraw-Hill, Cambridge, MA: MIT Press, 1996.
\bibitem{graphred} C.~P.~Wadsworth.
Semantics and Pragmatics of the Lambda Calculus, PhD thesis.
Oxford University, 1971.
\bibitem{kluge} W.~Kluge.
Abstract Computing Machines.
Springer-Verlag, 2005.
\bibitem{funcprog} A.~J.~Field, P.~G.~Harrison.
Functional Programming.
Addison-Wesley, 1988.
\bibitem{revesz} G.~R\'ev\'esz.
Axioms for the theory of lambda-conversion.
SIAM Journal on Computing, 14(2): 373--382, May 1985.
\bibitem{ustica} J.~W.~Klop. Term rewriting systems.
Notes prepared for the seminar on Reduction Machines,
organized by C. B\"ohm, Ustica, September 1985.
\bibitem{private} V.~van~Oostrom.
The inner spine strategy is normalising for
distributive $\lambda$-calculus.
Private correspondence.
\bibitem{explicit} K.~H.~Rose.
Explicit Substitution~--- Tutorial and Survey.
BRICS LS-96-3, 1996.
\bibitem{monograph} H.~P.~Barendregt.
The Lambda Calculus, Its Syntax and Semantics.
North-Holland, 1984.
\bibitem{heap} A.~Salikhmetov.
The Heap Lambda Machine.
arXiv:0806.4631v2, 2010.
\bibitem{selfinter} T.~\AE.~Mogensen.
Efficient self-interpretation in lambda calculus.
Journal of Functional Programming, 3(2): 345--363, 1992.
\bibitem{recreal} H.~G.~Rice.
Recursive Real Numbers.
Proceedings of the American Mathematical Society, 5(5):
784--791, October 1954.
\bibitem{strongred} R.~Hindley.
Axioms for strong reduction in combinatory logic.
The Journal of Symbolic Logic, 32: 224--236, 1967.
\bibitem{posix} The IEEE and The Open Group.
The Open Group Base Specifications Issue 6.
IEEE Standard 1003.1, 2004.
\bibitem{needred} H.~P.~Barendregt, J.~R.~Kennaway, J.~W.~Klop,
and M.~R.~Sleep.
Needed reduction and spine strategies for the lambda calculus.
Information and Computation, 75(3): 191--231, December 1987.
\bibitem{bloo} C.~J.~Bloo.
Preservation of Termination for Explicit Substitution.
PhD thesis, Tech\-nishche Universiteit Eindhoven, October 2, 1997.
\end{thebibliography}
